\documentclass[conference,letterpaper]{IEEEtran}

\usepackage{booktabs}
\usepackage{pifont}
\usepackage[T1]{fontenc}
\usepackage[utf8]{inputenc}
\usepackage{graphicx}
\usepackage{amsmath,amssymb}
\usepackage{cite}
\usepackage{url}
\usepackage[hidelinks]{hyperref}
\usepackage{comment}
\usepackage{tikz}
\usetikzlibrary{positioning,arrows.meta}

\newcommand{\etal}{\textit{et al.}}
\newcommand{\cmark}{\ding{51}}
\newcommand{\xmark}{\ding{55}}
\newcommand{\codelink}[1]{\href{#1}{\underline{Link}}}

\title{LLMs for Social Network Modeling: \\ From Network Generation to Dynamic Processes}

\newif\ifanonymous
\anonymousfalse

\ifanonymous
\author{\IEEEauthorblockN{Anonymous Authors}}
\else
\author{\IEEEauthorblockN{Shikha Mallick\IEEEauthorrefmark{1}, Alex Thomo\IEEEauthorrefmark{1}, and Akrati Saxena\IEEEauthorrefmark{2}}
\IEEEauthorblockA{\IEEEauthorrefmark{1}Department of Computer Science, University of Victoria, BC, Canada\\
\{shikhamallick, thomo\}@uvic.ca}
\IEEEauthorblockA{\IEEEauthorrefmark{2}Leiden Institute of Advanced Computer Science, Leiden University, The Netherlands\\
a.saxena@liacs.leidenuniv.nl}}
\fi

\begin{document}
\maketitle

\begin{abstract}
Large language models (LLMs) are rapidly emerging as a new paradigm for modeling social networks by representing users and their relationships and interactions through natural language. Unlike classical network models or deep learning approaches, LLMs can simulate context-aware social behavior and language-driven interactions, enabling more realistic modeling of network formation and dynamic social processes. However, existing studies are scattered across different research communities and lack a unified perspective. This survey presents the first comprehensive review of LLMs for social network modeling by organizing the literature into two broad categories: network generative models and dynamic process models. Network generative models are further classified into selection-based and interaction-based approaches, while dynamic process models are categorized into opinion dynamics, information diffusion, and rumor propagation, each with their underlying modeling mechanisms. LLMs enable rich textual social interactions and decision-making, but they also exhibit many limitations, including inherent social biases and prompt sensitivity. We outline these open research challenges and discuss future directions in LLM-based social network modeling.
\end{abstract}

\section{Introduction}

Social networks provide a natural framework for representing individuals, their relationships, and the interactions through which they communicate, exchange information, and influence one another \cite{newman2018networks}. In this context, social network analysis aims to understand the structure and dynamics of networks, that is, how connections, or edges, form between individuals and how opinions, information, and rumors evolve through networked interactions by simulating processes that are difficult to observe or manipulate in the real world.

Classical social network models use predefined rules to explain how networks evolve over time \cite{saxena2021evolving}. For example, network formation can be driven by preferential attachment, where well-connected nodes attract more connections \cite{barabasi1999emergence}, or by homophily, where individuals with similar attributes are more likely to connect \cite{saxena2025homophily}. Another example is that opinion change, information diffusion, and rumor spreading are modeled using predefined equations, probabilities, or thresholds that determine how users change their states or influence one another \cite{erdos1959random,watts1998collective,barabasi1999emergence,degroot1974reaching,kempe2003maximizing,daley1965stochastic,nekovee2007theory}. These models are efficient, interpretable, and theoretically grounded but rely on simplified user representations and predefined interaction rules. Deep learning methods provide a data-driven alternative by learning structural patterns directly from network data using deep graph generative models \cite{guo2022systematic,you2018graphrnn,bojchevski2018netgan,liao2019efficient}. Although they perform well across many network analysis tasks, they require task-specific training data, need to encode users and interactions as latent representations, and remain limited in modeling open-ended communication, context-dependent reasoning, and adaptive social behavior.

Recent advances in generative artificial intelligence (AI), such as Large Language Models (LLMs), have led to the development of autonomous agents that use LLMs to perceive their environment, retain context, reason about goals, and take actions with limited human intervention \cite{luo2025agent}. This has introduced a new paradigm for social network analysis by enabling these LLM agents to form networks and simulate social dynamics by exchanging messages, updating opinions, adopting behaviors, and propagating information or rumors through network interactions \cite{luo2025agent,de2023emergence,chang2025llms,chuang2024simulating,gao2023s3,hu2025rumor}. At the same time, LLM-based social network models introduce challenges in realism, scalability, bias, reproducibility, and validation against real-world behavior \cite{chang2025llms,mehdizadeh2025homophily,chuang2024simulating,luo2025agent}.

\begin{figure}[!t]
    \centering
    \includegraphics[width=\linewidth]{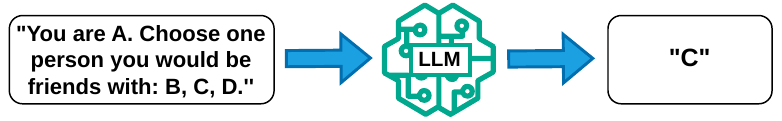}
    \caption{A simple example of prompting an LLM.}
    \label{fig:llm-prompt-example}
    \vspace{-2mm}
\end{figure}

Fundamentally, LLMs are pretrained natural language generative models built primarily on the Transformer architecture, which uses self-attention to model relationships among tokens in a textual sequence \cite{vaswani2017attention}. During pretraining, they learn statistical patterns from significantly large text corpora. Encoder-based models such as BERT learn contextual representations by reconstructing masked tokens \cite{devlin2019bert}, whereas autoregressive models such as GPT predict each token from its preceding context \cite{brown2020language}. Most LLMs used for agent simulation have billions of parameters, such as the GPT-3, Llama, Gemma, and Qwen families, and follow the autoregressive design because it supports open-ended generation, instruction following, and context-dependent responses \cite{zhao2026survey}.

LLMs can be adapted to a task through prompting or fine-tuning. A \emph{prompt} is an input text given to the LLM that specifies the task and relevant context \cite{zhao2026survey}. The LLM then processes the prompt and generates an output text as shown in Fig. \ref{fig:llm-prompt-example}. In role-based prompting, the LLM is given a persona or social role of a user with particular demographic characteristics, beliefs, or behavioral tendencies, and asked to form friendships \cite{chang2025llms}. Beyond task-specific prompt design, LLMs can also be fine-tuned for specific tasks using parameter-efficient approaches \cite{coppolillo2025engagement}. These capabilities make LLMs suitable for social network modeling, and existing studies have shown that repeated decisions among many LLM agents can produce population-level outcomes such as homophily, community formation, consensus, polarization, information cascades, and misinformation propagation \cite{de2023emergence,chang2025llms,chuang2024simulating,gao2023s3,hu2025rumor}.

\begin{figure}[!t]
    \centering
    \includegraphics[width=\linewidth]{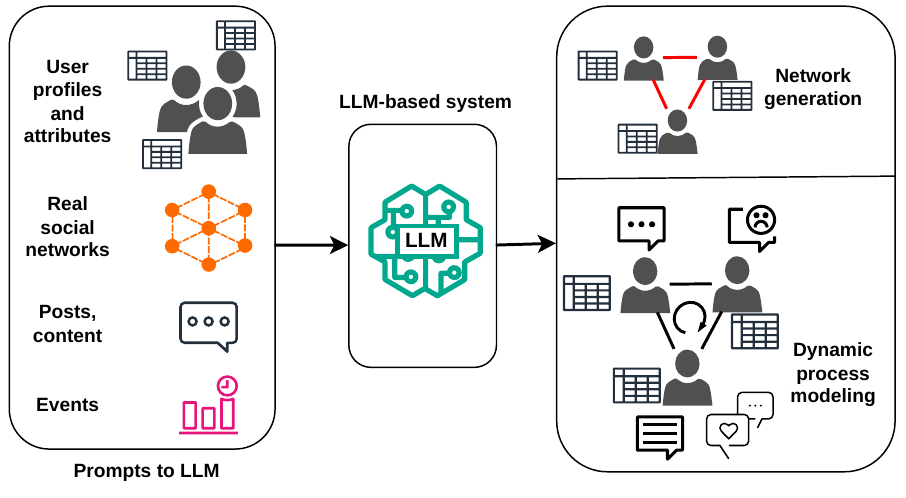}
    \caption{Overview of LLM-based social network modeling. User profiles, network structure, posts, and events are provided as textual context, and the LLM generates social connections or simulates dynamic processes among networked agents.}
    \label{fig:overview}
    \vspace{-2mm}
\end{figure}

As shown in Fig.~\ref{fig:overview}, prior works use LLMs in two principal ways. First, \emph{network generative models} infer or create edges from user attributes, structural information, or prior interactions. Second, \emph{dynamic process models} simulate how node states and actions change over time through networked interaction. Additionally, recent systems also incorporate evolving relationships, recommendation mechanisms, and platform interventions, creating feedback among users, network structure, and the online environment \cite{rossetti2024social,yang2024oasis,huang2026policysim}. This iterative use of LLMs differs from applying them only as classifiers or predictors to static network data. In an LLM-based social simulation, the LLM is part of the process being studied such that one agent's decision changes the information or network conditions encountered by other agents, whose later actions may feed back into the system.

Related surveys examine LLMs for graph learning, computational social science, social networks, text-attributed graphs, and general-purpose LLM agent systems \cite{jin2024large,li2023survey,thapa2025large,zeng2024large,su2025large,luo2025agent}. Broad agent surveys discuss profile construction, agent collaboration, and social network analysis as one of several application domains of agentic AI. However, they do not systematically organize how LLM agents generate social network structure or model the evolution of opinions, information, and rumors through networked interaction.

This survey addresses this gap through a unified, mechanism-centered review of LLM-based social network modeling. We organize existing methods into network generative models and dynamic process models, with the latter covering opinion dynamics, information diffusion, and rumor propagation. We systematically compare representative studies in terms of their LLM backbones, datasets, fine-tuning strategies, evaluation protocols, and code availability. Tables \ref{tab:supp-repro-part1} and \ref{tab:supp-repro-part2} mention the LLMs used for each of the different methods we discuss in this survey. Finally, we identify open challenges and future directions in LLM-based social network modeling.

\section{LLMs for Network Generative Models}
\label{sec:ngm}
Network generative models aim to generate synthetic networks that reproduce the topology and formation mechanisms of real-world social networks. LLMs can leverage linguistic and social knowledge acquired during pretraining for context-aware network formation \cite{zhao2026survey,chang2025llms,kilaru2026llms,de2023emergence}. Studies have shown LLM agents interacting and forming connections similar to how real social networks work. He \etal\ \cite{he2023homophily} analyzed Chirper.ai, a Twitter-like platform of autonomous LLM agents, and found homophilic communities emerging around shared language and semantically similar content. However, this platform offered limited control over agent design, prompting, and network-formation mechanisms, motivating controlled generators categorized by edge formation strategies. We categorize the controlled LLM-based network generation models based on the mechanism they adopt for network generation.

\begin{figure}[!t]
    \centering
    \includegraphics[width=\linewidth]{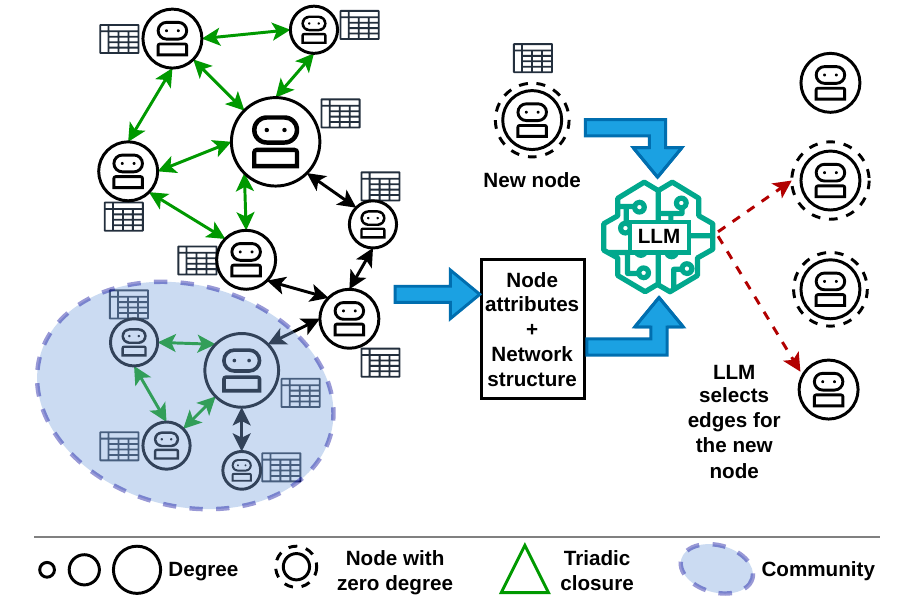}
    \caption{Basic mechanism of selection-based network generative models. LLM receives a new node profile, its attributes and the existing network structure and node attributes in the prompt. Then it selects the nodes with whom the new node should form edges.}
    \label{fig:llm-gen-sel}
    \vspace{-2mm}
\end{figure}

\subsection{Selection-based Network Generative Models}
In selection-based network generative models, the network may start empty or partially formed, but the number of nodes is always fixed. The LLM, while acting as a node in the network, directly selects nodes from the available set to form edges with. These models can be structure-conditioned or attribute-conditioned based on whether their generative process is conditioned on the network structure or node attributes. Fig. \ref{fig:llm-gen-sel} shows the basic mechanism of selection-based network generative models.

\paragraph{\textbf{Structure-conditioned}}
In these generative models, the decision is driven primarily by topological features, such as degree, neighborhoods, mutual neighbors, or community membership. De Marzo \etal\ \cite{de2023emergence} introduced one of the first LLM-based network generators in which the nodes enter an initially empty network sequentially. At each time step, a new node represented by an LLM agent receives the names and degrees of existing nodes, selects a fixed number of connection targets, and forms the corresponding edges. The authors found that repeated choices by the LLM produced heavy-tailed and approximately scale-free degree distributions; however, the LLM also favored certain names because of patterns in its training data, producing highly centralized networks.

Papachristou and Yuan \cite{papachristou2025network} conducted several controlled experiments that extend the sequential node-entry process of De Marzo \etal\ \cite{de2023emergence} through modified prompts. In their preferential-attachment experiment, the LLM agent receives either the degrees or complete neighborhood information of candidate targets. They found that complete neighborhood descriptions produced more realistic heavy-tailed structures than degree-only prompts, which often created highly centralized or star-like networks. Their other experiments study triadic closure, homophily, community formation, and small-world structure by providing the LLM with mutual-neighbor information, synthetic node attributes, or node information from real networks. These experiments revealed strong preferences for mutual neighbors and socially similar nodes.

GraphMind uses structure-aware supervised fine-tuning to infer missing edges from node profiles and serialized multi-hop paths \cite{bu2026beyond}. This approach improves structural fidelity by jointly leveraging node attributes and multi-hop structural information, although it may also inherit biases present in the training networks. The resulting realistic synthetic networks provide a useful benchmark for evaluating graph learning and bot detection algorithms.

\paragraph{\textbf{Attribute-conditioned}} 
In these generative models, the LLM agent's decision of adding edges is driven primarily by node attributes, such as demographics, interests, political affiliation, personality, culture, or language. To carefully examine what effect different node attributes have on the LLM's edge selections, Mehdizadeh and Hilbert \cite{mehdizadeh2025homophily} combined node popularity and attribute similarity with the sequential fixed-node network generation process, considering one categorical social attribute at a time, while node degree was always included as a structural feature. In their experiments, the LLM agent receives its attribute vector and the degrees and attribute vectors of existing nodes from a randomly selected subset of existing nodes. The resulting attachment probabilities are determined by the LLM’s stochastic weighting of attribute similarity, node degree, and the overall network structure. The authors found that the LLM favors a preferential attachment mechanism but substantially overestimates political and religious homophily.

Chang \etal\ \cite{chang2025llms} proposed three prompting strategies (Global/Local/Sequential) for generating networks from a fixed set of persona nodes whose demographic and political attributes were sampled from Census data. Global prompting gives all personas to the LLM at once and asks it to output the full edge set. Local prompting instead makes the LLM role-play each persona and select friends from the remaining nodes. Sequential prompting additionally provides the current friend lists or node degrees, so each choice depends on the partially generated network. Local and sequential prompting produce better connectivity, clustering, and degree heterogeneity, while sequential prompting captures long-tailed degree patterns more accurately. However, all three methods overestimate political homophily, and repeatedly including all personas in the prompts limits scalability.

Gkartzios \etal \cite{gkartzios2025modeling} examined how demographics, personality, reciprocity, and temporal dynamics shape the sequential fixed-node network generation process by varying which node attributes are revealed during connection decisions. In their experiments, a candidate pool of nodes for adding edges is constructed from the LLM agent's two-hop neighbors and a set of randomly sampled non-neighbors. Their results show strong demographic homophily and personality-related popularity effects, and other potential social biases in LLM behavior.

To carefully examine the effect of LLM edge selection beyond node attributes, Kilaru \etal\ \cite{kilaru2026llms} compared 192 directed networks generated over a fixed set of 50 personas across four cultures, four languages, three model scales, and four prompting methods. They used global, local, and sequential prompting methods \cite{chang2025llms}, although the latter two restrict the candidate set available to each persona. An additional iterative method repeatedly asks the LLM to revise an existing edge set. They found that these design choices substantially affected the generated networks. Smaller LLMs made qualitatively different, not merely noisier, friendship choices than larger LLMs, and although the generated networks had realistic clustering and modularity, political homophily was overestimated.

\begin{figure}[!t]
    \centering
    \includegraphics[width=\linewidth]{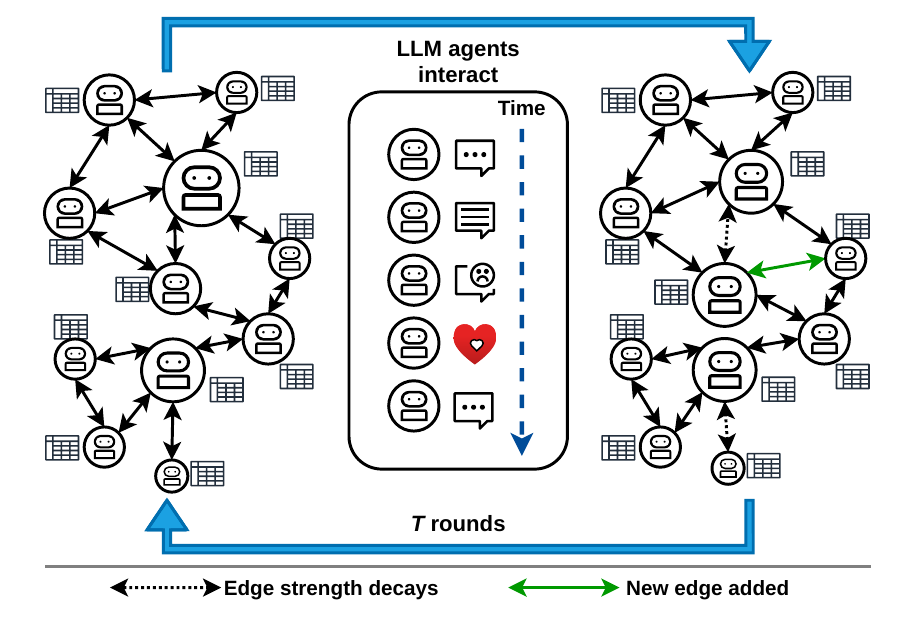}
    \caption{Basic mechanism of interaction-based network generative models. The nodes in the network are LLM agents who interact and form connections over $T$ rounds.}
    \label{fig:llm-gen-int}
    \vspace{-2mm}
\end{figure}

\subsection{Interaction-based Network Generative Models}
Interaction-based models derive social edges between nodes iteratively based on the repeated interaction of nodes over a fixed number of rounds. Fig. \ref{fig:llm-gen-int} shows the basic mechanism of interaction-based network generative models.

GraphAgent-Generator (GAG) is a multi-agent LLM framework that generates evolving actor-item bipartite networks to model how interactions between users and content produce different social networks over time \cite{ji2025llm}. Actor nodes represent authors, viewers, or social media users, while item nodes represent papers, movies, or posts. In each round, actors receive profiles, relevant items, and past interactions, then generate actions such as citing, rating or creating content. These actions form new edges, while new actors, items, and textual attributes are added over time. Projecting the generated bipartite networks yielded citation, co-authorship, recommendation and follow networks, demonstrating GAG's cross-domain flexibility. However, the framework does not systematically examine whether prompting produces the heterogeneous agent behaviors needed for realistic network formation.

Schneider \etal\ \cite{schneider2025learning} proposed a multi-agent LLM framework in which persistent LLM agents, equipped with personas, goals, memories, and social context, repeatedly communicate, evaluate others, and reflect on prior interactions. A heuristic or LLM-based update mechanism converts interaction histories into increasing or decaying relationship weights, allowing edges to emerge and weaken over time. This yields richer network trajectories but requires many LLM calls and careful calibration of rewards, update rules, and edge thresholds.

In summary, LLM-based network generating models differ in whether edges are selected directly or emerge based on prior behavior. They also vary depending on prompt design and LLM fine-tuning. Evaluation should therefore examine both final topology and the validity of the edge-formation process, including individual link choices and sensitivity to prompts, model families, and fine-tuning strategies. Reproducibility information for the reviewed network generative models is provided in Table \ref{tab:supp-repro-part1}.

\section{LLMs as Dynamic Process Models}
\label{sec:dpm}
Dynamic process models study how the states and actions of LLM agents change over time through networked interaction. Opinion dynamics is a representative example where each node begins with an initial opinion, and during the simulation, an LLM agent observes messages from neighboring nodes, interprets their arguments and context, and updates its own stance. Repeated interactions among agents produce consensus, fragmentation, or polarization. Related works study whether agents adopt or share information and whether they believe, modify, or retransmit rumors. We therefore organize LLM-based dynamic process models into opinion dynamics, information diffusion, and rumor propagation.

\begin{figure}[!t]
    \centering
    \includegraphics[width=\linewidth]{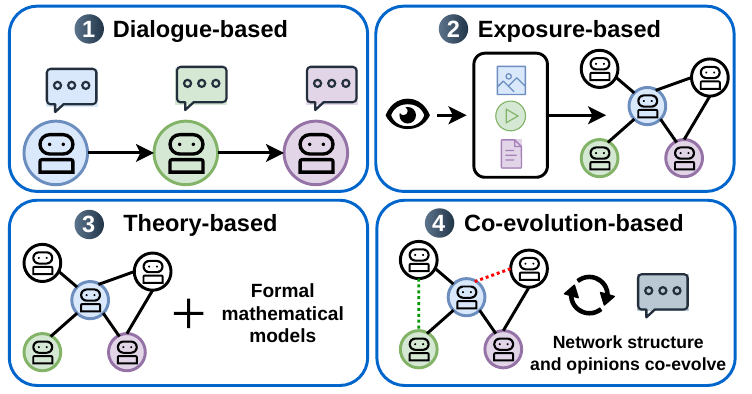}
    \caption{An overview of LLM-based opinion dynamics models.}
    \label{fig:llm-op-dyn}
    \vspace{-2mm}
\end{figure}

\subsection{Opinion Dynamics Models}
\label{ssec:odp}
Opinion dynamics studies how individual beliefs change through social influence and how these changes produce collective outcomes such as consensus, fragmentation, or polarization \cite{degroot1974reaching,friedkin1990social}. Networked LLM agents use initial stances, memory, and personas to generate messages and update opinions from conversational context. This captures persuasive content and heterogeneous responses, but may reproduce pretraining and alignment biases rather than realistic social behaviors \cite{chuang2024simulating,cisneros2025biases,yazici2026opinion}. Fig. \ref{fig:llm-op-dyn} shows the four types of LLM-based opinion dynamics models, classified by their modeling mechanisms.

\subsubsection{Dialogue-based Opinion Dynamics Models}
Dialogue-based models update node opinions through conversational reasoning, memory, or visible discussion histories. Chuang \etal \cite{chuang2024simulating} introduced a population of networked LLM agents that interact and repeatedly update their beliefs. The agents converged toward fact-aligned consensus without an induced cognitive bias. In contrast, prompted confirmation bias produced persistent disagreement and fragmented opinion clusters. This established both the promise of language-based simulation and a central limitation that realistic disagreement may need to be explicitly prompted. Cisneros-Velarde \cite{cisneros2025biases} removed the dependence on explicitly prompted biases and personas. Instead, they examined built-in LLM tendencies during controlled funding discussions by varying the ethical connotations, memory, and the format of opinion updates (open-ended versus multiple-choice). Cau \etal \cite{cau2025selective} analyzed multi-round LLM debates and found selective persuasion rather than indiscriminate agreement. LLM agents are more receptive to positions closer to their existing views, and logically fallacious arguments can still influence opinion change. Zhong \etal \cite{zhong2025spiral} studied spiral-of-silence dynamics through sequential movie ratings rather than agent-to-agent dialogue. They observed that persona alone preserves diversity, history alone creates anchoring, and combining both produces self-reinforcing majority dominance.

\subsubsection{Exposure-based Opinion Dynamics Models}
Exposure-based models study how network topology, recommendations, selective attention, or manipulated communication determines the information available to each node, thereby shaping collective opinion. Ferraro \etal\ \cite{ferraro2024agent} simulated LLM-based social-network users and examined how personalized recommendations affect engagement, homophily, and echo-chamber formation. Ohagi \cite{ohagi2024polarization} focused on polarization under echo-chamber exposure, in which LLM agents mainly interacted with others holding similar views, which drove their opinions toward extremes. Exposure to a wider range of viewpoints more often produced convergence; however, the effect also depended on persona design and the initial opinion distribution. Wang \etal \cite{wang2025decoding} introduced recommendation-mediated exposure over alternative network structures using LLM agents. LLM agents reason over neighboring messages before updating their positions. The framework reproduces echo chambers more effectively than numerical baselines and supports language-based prompts for mitigation. Donkers and Ziegler \cite{donkers2025understanding} extended evaluation to humans embedded in an LLM-populated social network platform. Their controlled study shows that polarized environments heighten emotionality and group-identity salience while reducing expressed uncertainty. Zheng and Tang \cite{zheng2026simulating} extended this platform view by deploying LLM agents that post and react on small-world and scale-free networks, showing that network topology shapes interaction patterns and echo-chamber formation. Dehkordi \etal \cite{dehkordi2026opinion} show that a small set of adversarial LLM agents can increase polarization through natural-language manipulation. The tested defenses reduce but do not eliminate this effect. Finally, Liu \etal \cite{liu2026social} distinguished visible exposure from realized influence and showed that selective attention can move LLM populations from collective intelligence toward herding.

\subsubsection{Theory-based Opinion Dynamics Models}
Theory-based models constrain, compare, or interpret LLM-generated opinion trajectories using formal opinion dynamics mechanisms or empirically calibrated behavioral models. An early hybrid formulation by Li \etal \cite{li2023quantifying} incorporated LLM-generated opinions into a mathematical model of collective opinion dynamics. Mou \etal\ \cite{mou2024unveiling} proposed a hybrid framework in which LLM-based core users interact with ordinary users governed by formal opinion-dynamics models. Using real Twitter networks and social-movement data, the framework evaluates both individual behavior alignment and population-level attitude trajectories. Yazici \etal \cite{yazici2026opinion} showed that networked LLMs often reach consensus at a rate related to network structure. However, the final position differs from the DeGroot prediction \cite{degroot1974reaching} and depends strongly on topic-specific model bias. He \etal \cite{he2026opinion} implemented DeGroot-like \cite{degroot1974reaching} and Friedkin-Johnsen-like \cite{friedkin1990social} dialogues more directly and identified substantial differences in self-trust and susceptibility across LLM families. Lan \etal \cite{lan2026public} combine probabilistically calibrated behavior profiles with LLM-based cognition to reproduce broader public-opinion trajectories. Probine \etal \cite{probine2026characterizing} fit interpretable opinion models to LLM trajectories and find that innate bias and social averaging explain more variation than stubbornness or homophily.

\subsubsection{Co-evolution-based Opinion Dynamics Models}
More recent work relaxes the assumption that the communication environment remains fixed. Co-evolution-based models allow opinions to evolve jointly with the communication structure or semantic environment, creating reciprocal feedback between LLM agents' beliefs and the interactions or topics that subsequently influence them. Piao \etal \cite{piao2025emergence} allowed LLM agents to form social edges while their political opinions evolved, resulting in homophilic clusters and echo-chamber-driven polarization. Gu \etal \cite{gu2025large} modeled this feedback more explicitly by allowing LLM agents to update both their opinions and their connections. Composta \etal \cite{composta2026simulating} calibrated LLM agent profiles and initial opinions using real social-media data and examined the joint evolution of conversations, beliefs, and relationships. SPARK further extended opinion dynamics to semantic evolution, showing that stance shifts affect which subtopics emerge and that these subtopics subsequently reshape LLM agent stances \cite{zhang2025spark}.

At a broader level, the main advantage of networked LLM agents for modeling opinion dynamics is their ability to represent the content and framing of social influence. Their central limitation is that opinion updating is delegated to an opaque and model-dependent mechanism. Evaluation should therefore combine trajectory-level and population-level measures with robustness tests and validation against human behavior. The information to reproduce these methods is summarized in Tables \ref{tab:supp-repro-part1} and \ref{tab:supp-repro-part2}.

\begin{figure}[!t]
    \centering
    \includegraphics[width=\linewidth]{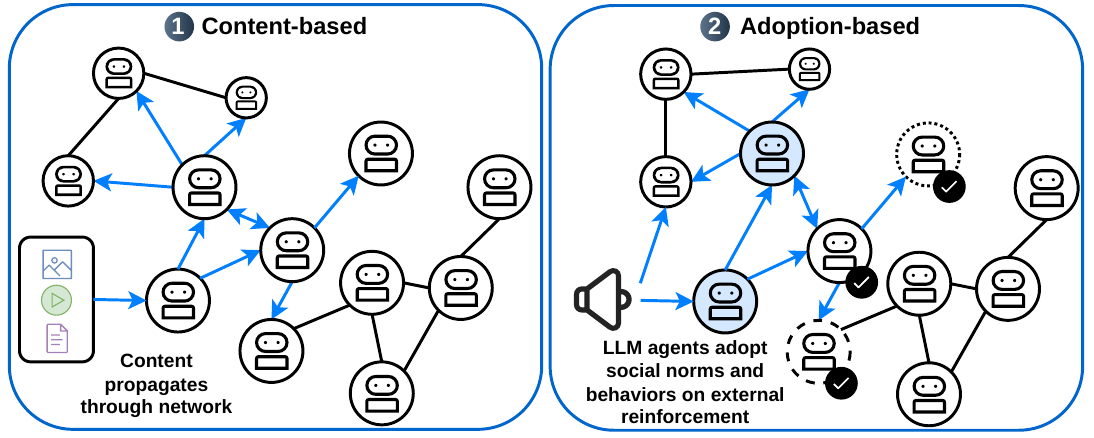}
    \caption{The two categories of LLM-based information diffusion models: content-based and adoption-based.}
    \label{fig:llm-inf-diff}
    \vspace{-2mm}
\end{figure}

\subsection{Information Diffusion Models}
\label{ssec:iibd}
Information diffusion studies how information spreads through a social network over time. The process begins with a small set of seed nodes that adopt or share information, changing their state from inactive to active. Their neighbors are then exposed and may also adopt or share it, while newly activated nodes subsequently expose their own neighbors. This process continues until no further nodes become active or the simulation ends \cite{goldenberg2001talk,kempe2003maximizing}. LLM-based information diffusion models introduce language-aware actors that can interpret content and generate context-dependent reactions. Fig. \ref{fig:llm-inf-diff} shows two types of LLM-based information diffusion models, distinguished by their modeling mechanisms.

\subsubsection{Content-based Information Diffusion Models}
Content-based models simulate the propagation of posts, statements, and other content through a network. These models differ in whether the LLM is an individual user, a content generator, or part of an aggregate population group. Gao \etal\ \cite{gao2023s3} proposed \(S^3\), which grounds LLM agents in real social-network data and equips them with demographic profiles and interaction histories. After observing posts from followed users, agents may create content, forward existing content, or remain inactive, producing population-level diffusion trajectories and changes in attitudes and emotions. However, individual-level simulation incurs substantial computational cost. Coppolillo \etal\ \cite{coppolillo2025engagement} proposed an optimized method for content reach in which an LLM-controlled node generates a post to engage other nodes in the network, while a formal engagement model simulates the cascade and uses the number of activated nodes as a reinforcement-learning reward. This enables adaptation to the network's opinion distribution and topology, but the results depend on the assumptions of the proxy diffusion model. LLM-AIDSim combines linguistic responses with classical influence diffusion \cite{zhang2025llm}. It retains an Independent-Cascade-style activation process \cite{goldenberg2001talk,kempe2003maximizing}, in which activated agents generate reactions to the original statement and to neighboring responses. It captures discourse evolution beyond cascade size, although activation and response updates still rely on externally specified probabilities. GA-\(S^3\) improves scalability by replacing individual users with hierarchical group agents \cite{zhang2025ga}, where each group represents users with similar attributes. This supports much larger effective populations at lower cost, but may hide within-group differences.

\subsubsection{Adoption-based Information Diffusion Models}
Adoption-based methods examine how LLM agents as nodes adopt behaviors, policies, technologies, cooperative actions, or social norms through social reinforcement. Hitz \etal \cite{hitz2025amplifier} found that LLM agents exhibit lower adoption thresholds than human participants, causing policies and technologies to spread more widely in mixed human-agent populations. PROSIM examined the evolution of prosocial actions among LLM agents under policy-induced inequality and showed that unfair interventions can suppress cooperation and propagate norm erosion through the network \cite{zhou2025simulating}.

On the whole, these studies extend information diffusion modeling from the circulation of textual content to the adoption of behaviors and norms. Reproducibility information for these models is reported in Table \ref{tab:supp-repro-part2}.

\begin{figure}[!t]
    \centering
    \includegraphics[width=\linewidth]{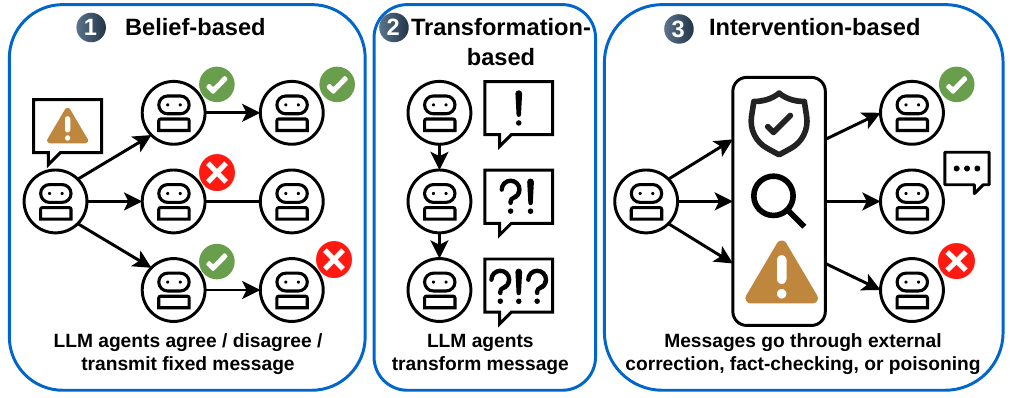}
    \caption{The three categories of LLM-based rumor propagation models: belief-based, transformation-based, and intervention-based.}
    \label{fig:llm-rum-prop}
    \vspace{-2mm}
\end{figure}

\subsection{Rumor Propagation Models}
\label{ssec:rmp}
Rumor propagation concerns not only how widely content spreads, but also whether users believe it and how its meaning changes during transmission \cite{saxena2021fake}. LLM agents provide a language-aware mechanism by interpreting the message before deciding whether to believe, modify, or retransmit it. Fig. \ref{fig:llm-rum-prop} shows the three types of LLM-based rumor propagation models based on their modeling mechanisms.

\subsubsection{Belief-based Rumor Propagation Models}
Belief-based methods assume that a rumor or false claim already exists and simulate whether agents accept, reject, discuss, or retransmit it. Liu \etal \cite{liu2024skepticism} first proposed FPS, where persona-conditioned LLM agents exchange opinions about pre-existing fake news and update their beliefs through memory and reflection. The framework reproduces topic-dependent and personality-dependent susceptibility and shows that early repeated correction is more effective than delayed intervention. Li \etal \cite{li2026newsdiffusion} subsequently replaced fixed forwarding probabilities with LLM decisions conditioned on psychological traits. Their experiments show that both cascade size and intervention effectiveness depend on network structure. Hu \etal \cite{hu2025rumor} studied decentralized rumor spreading across synthetic and real social networks. Each LLM agent maintains beliefs and a posting history, while persona prompts control its tendency to accept and forward rumors. RumorSphere enabled a million-agent simulation through a reinforced hybrid architecture \cite{liu2026rumorsphere} by dynamically assigning LLM reasoning to agents near boundaries between opposing belief groups and modeling the remaining population with efficient behavioral rules.

\subsubsection{Transformation-based Rumor Propagation Models}
Transformation-based methods model how message content changes during transmission, including semantic distortion, claim loss, and factual degradation. FUSE focused on the gradual distortion of initially true news \cite{liu2025stepwise}. Spreaders, commentators, and verifiers alter content in different ways, while FUSE-EVAL measures its deviation from the source. Maurya \etal \cite{maurya2025misinformation} focused on factual degradation along persona-conditioned transmission chains. A question-answering (QA)-based auditor tracks which claims survive each rewrite.

\subsubsection{Intervention-based Rumor Propagation Models}
Intervention-based methods examine malicious or corrective actions, including content poisoning, censorship, reporting, and fact-checking. TrendSim modeled poisoning attacks on centrally displayed trending topics \cite{zhang2025trendsim}. Its time-aware LLM agents interact with malicious comments and support experiments on attacker prevalence and content censorship. MOSAIC compared community-based, third-party, and hybrid fact-checking within a directed social graph \cite{liu2025mosaic} where safety-aligned LLMs appeared less susceptible to rumors than human users.

Collectively, these methods move beyond whether rumors are believed and shared to examine semantic transformation, malicious amplification, fact-checking, and propagation in large mixed populations. Evaluation should therefore measure reach, belief changes, factual loss, moderation success, and human realism. Table~\ref{tab:supp-repro-part2} in the supplementary material summarizes the LLM backbones, fine-tuning choices, datasets, and code availability of the reviewed rumor propagation models.

\section{Open Challenges and Future Directions}
\label{sec:challenges}
In this section, we describe several open challenges in LLM-based social network models identified through our literature survey, and provide suggestions for future work for making these models robust.

\subsection{Validation, Benchmarking, and Reproducibility}
It remains a challenge to determine whether LLM-agent behavior is realistic, whether simulated mechanisms explain social outcomes, and whether results can be reproduced across studies. Agents may produce plausible individual decisions while generating unrealistic collective patterns such as excessive homophily, consensus, or polarization \cite{chang2025llms,chuang2024simulating}. Human-agent experiments, calibrated benchmarks, and policy or feed interventions provide initial validation approaches \cite{donkers2025understanding,zhou2025simulating,tornberg2023simulating,zhang2026intervensim,huang2026policysim}, but behavioral similarity does not establish causal validity \cite{pearl2009causality,goyal2026plausible}. Results can also vary with the LLM, prompt, and simulation configuration \cite{chang2025llms,kilaru2026llms,luo2025agent}. Future benchmarks should combine real network structures with time-stamped interaction data, validate individual and network-level outcomes, and report prompts, update rules, computational costs, and simulation settings for reproducibility.

\subsection{Scalable Simulation}
Repeatedly querying an LLM for every agent across many interaction rounds makes large-scale simulations computationally expensive \cite{luo2025agent}. Recent works such as OASIS use distributed simulation, GA-\(S^3\) represents similar users through group agents, and RumorSphere combines selective LLM reasoning with lightweight behavioral rules, enabling million-agent experiments \cite{yang2024oasis,zhang2025ga,liu2026rumorsphere}. However, group-level aggregation may obscure within-group behavioral heterogeneity \cite{zhang2025ga,zhang2026intervensim}. Multi-fidelity simulators that invoke full LLM reasoning only for complex decisions while quantifying the trade-off between computational cost and population representativeness are needed. 

\subsection{Fairness, Bias, and Responsible Simulation}
Fairness is important in social network analysis because biases
in networks can affect downstream tasks such as link
prediction, influence maximization, centrality ranking, and
community detection \cite{saxena2024fairsna}. In LLM-based social network models, biased agent decisions can similarly distort connectivity, exposure, and influence \cite{chang2025llms,mehdizadeh2025homophily,tornberg2023simulating}. Existing LLM-based studies report exaggerated homophily \cite{chang2025llms,mehdizadeh2025homophily}, while other work examines cross-group feeds \cite{tornberg2023simulating} and fairness-aware influence maximization \cite{wang2025fairim}. Fairness in opinion dynamics has also been studied in conventional network models \cite{stepien2026fairness}, providing a useful direction for future LLM-based simulations. Another promising direction is to extend topic-specific influential-user detection \cite{panchendrarajan2023topic} by combining network measures with LLM-based content understanding. Constructing agents using real data also raises privacy concerns, while optimizing content reach may introduce risks related to persuasion and influence \cite{thapa2025large,yang2024oasis,coppolillo2025engagement,luo2025agent}. Future work should develop fairness-aware models for network formation, opinion dynamics, and influence analysis, while protecting privacy, reporting uncertainty, and guarding against harmful uses without sacrificing realistic behavior and network structure.

\subsection{Dynamic and Co-evolving Networks} Most methods capture only part of the feedback among social ties, user states, content, and platform exposure. Recent models allow opinions and connections to co-evolve or combine persistent agents with changing social environments \cite{piao2025emergence,gu2025large,composta2026simulating,rossetti2024social}. However, maintaining reliable behavior over long simulations remains difficult, and current frameworks capture only selected forms of temporal change \cite{luo2025agent,yang2024oasis}. Hence, methods that jointly model tie formation and dissolution, evolving identities and memories, and platform-mediated exposure within a unified temporal framework are needed.

\subsection{Theory-Guided Foundation Models} Most current systems rely on general-purpose LLMs whose simulated social behavior may reflect model-specific biases rather than explicit social or network mechanisms \cite{chuang2024simulating,cisneros2025biases,yazici2026opinion}. Their behavioral update rules also remain opaque and vary across topics and model families, limiting interpretability, reproducibility, and transfer \cite{yazici2026opinion,he2026opinion,probine2026characterizing}. Recent hybrid approaches combine formal social-dynamics models with LLMs to provide testable mechanisms while capturing the semantic content and context of social interaction \cite{li2023quantifying,mou2024unveiling,lan2026public}. However, current theory-guided approaches remain largely task-specific and concentrated on opinion dynamics. Future work should develop social-network foundation models trained on multimodal interaction data and constrained by network science, behavioral evidence, and causal social theories.

\subsection{Hybrid Human-LLM Social Networks}
Future social networks are likely to include both human and LLM-based agents that interact and influence one another \cite{pedreschi2023human,han2026socialphysics}. Early studies suggest that hybrid networks may exhibit collective behaviors distinct from purely human or AI systems \cite{gaubert2026experimental}. Existing network models may not fully capture heterogeneous human-AI interactions \cite{pedreschi2023human,han2026socialphysics}. Future work should develop methods to analyze human-AI network formation, asymmetric influence, evolving social roles, and the coevolution of network structure and behavior. 

\section{Conclusion}
The development of generative AI, including LLMs and agentic AI, has made social network analysis and modeling an active research area across a growing range of applications. This survey provides a unified, mechanism-centered review of LLM-based social network modeling. We organized prior work into network generative models, including selection-based and interaction-based approaches, and dynamic process models covering opinion dynamics, information diffusion, and rumor propagation. Across these areas, LLM agents enable richer representations of users, communication, and context-dependent social behavior than classical rule-based models. However, current methods face limitations in behavioral realism, scalability, bias, reproducibility, and validation against real social systems. We highlighted future directions involving better benchmarks, scalable simulation, dynamic and co-evolving networks, fairness-aware modeling, and theory-guided approaches. As human and LLM agents increasingly interact online, understanding the evolution of hybrid human-LLM networks may also become important. Overall, progress will likely depend on combining LLM capabilities with rigorous network science, behavioral validation, and responsible simulation practices.

\clearpage

\bibliographystyle{IEEEtran}
\bibliography{references}

\clearpage

\setcounter{table}{0}
\renewcommand{\thetable}{S\arabic{table}}

\begin{table*}[!t]
\centering
\begin{minipage}{\textwidth}
\normalsize
\section*{Supplementary Material}
This supplement reports reproducibility information for the studies reviewed in the survey. The tables preserve the reported LLM backbones, fine-tuning choices, real datasets, and public code availability. Boldface identifies either the sole task LLM evaluated in a study or the best-performing LLM explicitly identified by the authors. No boldface indicates that no unique aggregate winner was reported or that model rankings depend on the evaluation metric or experimental condition. \textit{FT} denotes fine-tuning. $^{*}$ denotes parameter-efficient fine-tuning, such as LoRA or prefix tuning. \text{ }\cmark\text{ } indicates that fine-tuning was used, whereas \text{ }\xmark\text{ } indicates that it was not. ``Optional'' means that the framework supports configurations with and without fine-tuning. -- indicates that no applicable real dataset was used or that public code availability was not reported.
\end{minipage}
\vspace{0.5mm}
\caption{Reproducibility Information for Reviewed Models (Part I)}
\label{tab:supp-repro-part1}
\footnotesize
\renewcommand{\arraystretch}{1.00}
\setlength{\tabcolsep}{3pt}
\begin{tabular}{p{2.8cm} p{4.45cm} c p{7.30cm} c}
\toprule
\textbf{Method} &
\textbf{LLM(s)} &
\textbf{FT} &
\textbf{Real Datasets Used} &
\textbf{Code} \\
\midrule

\multicolumn{5}{l}{\textbf{Network Generative Models: Selection-based, Structure-conditioned}} \\
\midrule
De Marzo \etal\ \cite{de2023emergence} &
\textbf{GPT-3.5-turbo} &
\xmark &
-- &
-- \\

Papachristou-Yuan \cite{papachristou2025network} &
GPT-3.5-turbo; GPT-4-1106-preview &
\xmark &
Facebook100, including UChicago30 and Swarthmore42 &
\codelink{https://github.com/papachristoumarios/llm-network-formation} \\

Bu \etal\ (GraphMind) \cite{bu2026beyond} &
Qwen3-1.7B; Qwen3-7B &
\cmark$^{*}$ &
Cresci-15; Cresci-17; TwiBot-20; TwiBot-22; MGTAB-22 &
-- \\

\midrule
\multicolumn{5}{l}{\textbf{Network Generative Models: Selection-based, Attribute-conditioned}} \\
\midrule
Mehdizadeh-Hilbert \cite{mehdizadeh2025homophily} &
Gemini 1.5 Flash; GPT-4o Mini; Claude 3 Haiku; Llama-4-Scout &
\xmark &
-- &
\codelink{https://github.com/AliakbarMehdizadeh/llm-network-homophily} \\

Chang \etal\ \cite{chang2025llms} &
\textbf{GPT-3.5 Turbo}; GPT-4o; Llama 3.1 (8B/70B); Gemma 2 (9B/27B) &
\xmark &
U.S. Census-derived personas; eight CASOS/KONECT friendship networks; MySpace; Tuenti; GSS homophily statistics &
\codelink{https://github.com/snap-stanford/llm-social-network} \\

Gkartzios \etal\ \cite{gkartzios2025modeling} &
\textbf{Qwen3-30B} &
\xmark &
U.S. Census ACS demographic categories and population marginals &
-- \\

Kilaru \etal\ \cite{kilaru2026llms} &
GPT-4.1; GPT-4.1-mini; GPT-4.1-nano &
\xmark &
U.S. Census-derived persona distributions &
-- \\

\midrule
\multicolumn{5}{l}{\textbf{Network Generative Models: Interaction-based}} \\
\midrule
Ji \etal\ (GAG) \cite{ji2025llm} &
GPT-3.5-turbo; GPT-4o-mini; Llama-3-70B; Qwen2-72B &
\xmark &
Cora; CiteSeer; MovieLens &
\codelink{https://github.com/Ji-Cather/GraphAgent} \\

Schneider \etal\ \cite{schneider2025learning} &
\textbf{GPT-4o mini} &
\xmark &
Empirical social-network statistics reported in prior studies; no raw real-network dataset &
-- \\

\midrule
\multicolumn{5}{l}{\textbf{Opinion Dynamics Models: Dialogue-based}} \\
\midrule
Chuang \etal\ \cite{chuang2024simulating} &
ChatGPT; GPT-4; Vicuna-33B-v1.3 &
\xmark &
Climate-change statements and empirically grounded opinion settings &
\codelink{https://github.com/yunshiuan/llm-agent-opinion-dynamics} \\

Cisneros-Velarde \cite{cisneros2025biases} &
Llama 3; Mistral &
\xmark &
-- &
\codelink{https://drive.google.com/file/d/1-tVImMkThhBhWOLBcvAr3FZ49OCQOLZQ/view} \\

Cau \etal\ \cite{cau2025selective} &
Mistral-7B-Instruct-v0.2; Llama-3-8B-Instruct &
\xmark &
-- &
\codelink{https://github.com/ericacau/LLM-Opinion-Dynamics} \\

Zhong \etal\ \cite{zhong2025spiral} &
GPT-4o-mini; DeepSeek-V2-Lite-Chat; Mistral-8B-Instruct; Qwen2.5 (1.5B/3B/7B) &
\xmark &
IMDb films released after January 2025; PersonaHub personas &
\codelink{https://github.com/aialt/SoS-LLMs} \\

\midrule
\multicolumn{5}{l}{\textbf{Opinion Dynamics Models: Exposure-based}} \\
\midrule
Ohagi \cite{ohagi2024polarization} &
GPT-3.5-turbo-0613; GPT-4-0613 &
\xmark &
-- &
-- \\

Wang \etal\ \cite{wang2025decoding} &
\textbf{GPT-4o-mini} &
\xmark &
-- &
\codelink{https://github.com/ZongfangLiu/EchoChamberSim} \\

Donkers-Ziegler \cite{donkers2025understanding} &
\textbf{GPT-4o-mini} &
\xmark &
Responses and behavioral data from 122 human participants &
-- \\

Zheng-Tang \cite{zheng2026simulating} &
\textbf{GPT-4o} &
\xmark &
Roe v.\ Wade case materials; synthetically generated networks &
-- \\

Dehkordi \etal\ \cite{dehkordi2026opinion} &
\textbf{GPT-4.1-mini}; GPT-4o-mini; DeepSeek &
\xmark &
Twitter and Reddit opinion-network datasets &
\codelink{https://github.com/aSafarpoor/AA} \\

Liu \etal\ (SNLA) \cite{liu2026social} &
Qwen2.5-7B; Llama-3.1-8B &
\xmark &
-- &
-- \\

\midrule

\multicolumn{5}{l}{\textbf{Opinion Dynamics Models: Theory-based}} \\
\midrule
Li \etal\ \cite{li2023quantifying} &
LLM-generated opinions; backbone not specified &
\xmark &
-- &
-- \\

Mou \etal\ (SoMoSiMu) \cite{mou2024unveiling} &
\textbf{GPT-3.5-Turbo-0613} &
\xmark &
Metoo, RoeOverturned, and BlackLivesMatter Twitter datasets &
\codelink{https://github.com/xymou/social_simulation} \\

Yazici \etal\ \cite{yazici2026opinion} &
\textbf{Gemini 2.0 Flash}; GPT-5-nano for scoring and ablations &
\xmark &
-- &
\codelink{https://huggingface.co/datasets/asl-epfl/Social-LLM-Networks} \\

He \etal\ \cite{he2026opinion} &
\textbf{DeepSeek V3}; Qwen2.5; Mistral Large; GPT-4o mini; Llama 3.3 &
\xmark &
-- &
\codelink{https://github.com/hreyulog/llm_opinion_dynamic} \\

Lan \etal\ \cite{lan2026public} &
Single LLM backbone; model name not reported &
\xmark &
Weibo public-policy and food-safety event data &
-- \\

Probine \etal\ \cite{probine2026characterizing} &
Llama-3.1-8B; Qwen3-VL-8B-Instruct; Gemma-3-4B-it &
\xmark &
-- &
-- \\

\bottomrule
\end{tabular}
\end{table*}

\clearpage

\begin{table*}[!t]
\centering
\caption{Reproducibility Information for Reviewed Models (Part II)}
\label{tab:supp-repro-part2}
\footnotesize
\renewcommand{\arraystretch}{1.00}
\setlength{\tabcolsep}{3pt}
\begin{tabular}{p{2.8cm} p{4.45cm} c p{7.30cm} c}
\toprule
\textbf{Method} &
\textbf{LLM(s)} &
\textbf{FT} &
\textbf{Real Datasets Used} &
\textbf{Code} \\

\midrule

\multicolumn{5}{l}{\textbf{Opinion Dynamics Models: Co-evolution-based}} \\
\midrule
Piao \etal\ \cite{piao2025emergence} &
GPT-3.5 Turbo; GPT-4o; ChatGLM; Llama-3 &
\xmark &
-- &
-- \\

Gu \etal\ \cite{gu2025large} &
ChatGPT; GPT-4o Mini; Gemini; Gemma2-27B; Llama3.1-70B; Qwen2-72B &
\xmark &
COVID-19 vaccine and Russia-Ukraine War Twitter datasets &
-- \\

Composta \etal\ \cite{composta2026simulating} &
Llama-2-70B-Uncensored; Llama-3.2-3B-Uncensored &
\xmark &
ITA-ELECTION-2022; Bluesky-calibrated activity; X/Twitter demographic statistics &
\codelink{https://github.com/elisacomposta/YAnalysis} \\

Zhang \etal\ (SPARK) \cite{zhang2025spark} &
\textbf{DeepSeek-V3-0324} &
\xmark &
Five real-world discussion domains; no direct social-network dataset used &
\codelink{https://github.com/yangyi626/SPARK_} \\

\midrule
\multicolumn{5}{l}{\textbf{Information Diffusion Models: Content-based}} \\
\midrule
Gao \etal\ ($S^3$) \cite{gao2023s3} &
GPT-3.5; ChatGLM-6B &
Optional &
Weibo social-network, event, and user-interaction data &
-- \\

Coppolillo \etal\ \cite{coppolillo2025engagement} &
\textbf{Gemma-2B}; Mistral-7B; Llama variants; GPT-2; ChatGPT-4o &
\cmark &
Brexit and 2016 Italian constitutional-referendum X/Twitter networks &
\codelink{https://github.com/mminici/Engagement-Driven-Content-Generation} \\

Zhang \etal\ (LLM-AIDSim) \cite{zhang2025llm} &
\textbf{Llama 3} &
\cmark &
New York Times Articles and Comments &
-- \\

Zhang \etal\ (GA-$S^3$) \cite{zhang2025ga} &
\textbf{LLaMA3-8B}; Kimi and GPT-4 for auxiliary processing &
\xmark &
Social Network Benchmark with 30 online events from 2024 &
\codelink{https://github.com/AI4SS/GAS-3} \\

\midrule
\multicolumn{5}{l}{\textbf{Information Diffusion Models: Adoption-based}} \\
\midrule
Hitz \etal\ \cite{hitz2025amplifier} &
GPT-3.5-turbo; Gemini-1.5-Flash &
\xmark &
Human conjoint experiments on policy and technology adoption; Add Health social network &
-- \\

Zhou \etal\ (PROSIM) \cite{zhou2025simulating} &
LLaMA-3-8B; Qwen2.5-7B; DeepSeek-V3; GPT-3.5-turbo; \textbf{GPT-4o} &
\xmark &
National demographic distributions; 104-participant human benchmark &
\codelink{https://github.com/halsayxi/ProSim/} \\

\midrule
\multicolumn{5}{l}{\textbf{Rumor Propagation Models: Belief-based}} \\
\midrule
Liu \etal\ (FPS) \cite{liu2024skepticism} &
\textbf{GPT-3.5-turbo-1106} &
\xmark &
Real fake-news cases: political, terrorism, disaster, science, financial, and urban-legend topics &
\codelink{https://github.com/LiuYuHan31/FPS} \\

Li \etal\ \cite{li2026newsdiffusion} &
\textbf{GPT-3.5-turbo-1106} &
\xmark &
FakeNewsNet; empirical Twitter community derived from the Higgs dataset &
-- \\

Hu \etal\ \cite{hu2025rumor} &
\textbf{GPT-4o-mini} for agents; GPT-4 for persona generation &
\xmark &
Facebook \#686 network &
\codelink{https://github.com/UT-SysML/rumors-in-multi-agent} \\

Liu \etal\ (RumorSphere) \cite{liu2026rumorsphere} &
\textbf{GPT-4o-mini} &
\xmark &
Moon Landing Conspiracy; Xinjiang Cotton; Trump-Russia Connection Twitter datasets &
-- \\

\midrule
\multicolumn{5}{l}{\textbf{Rumor Propagation Models: Transformation-based}} \\
\midrule
Liu \etal\ (FUSE) \cite{liu2025stepwise} &
GPT-4o-mini; GPT-4 &
\xmark &
120 real news articles across five topic domains &
\codelink{https://github.com/LiuYuHan31/FUSE} \\

Maurya \etal\ \cite{maurya2025misinformation} &
\textbf{GPT-4o} &
\xmark &
Online news articles from real media sources across ten domains &
\codelink{https://github.com/RajGM/LLM-backend/} \\

\midrule
\multicolumn{5}{l}{\textbf{Rumor Propagation Models: Intervention-based}} \\
\midrule
Zhang \etal\ (TrendSim) \cite{zhang2025trendsim} &
\textbf{GLM-3-turbo} &
\xmark &
1,000 anonymized social-media user profiles; 10 trending topics &
\codelink{https://github.com/nuster1128/TrendSim} \\

Liu \etal\ (MOSAIC) \cite{liu2025mosaic} &
GPT-4o; DeepSeek-V3; Claude-3.7-Sonnet &
\xmark &
NewsGuard articles; personas derived from 204 Prolific participants &
\codelink{https://github.com/genglinliu/MOSAIC} \\
\bottomrule
\end{tabular}
\end{table*}

\end{document}